\documentclass[
 prx,
 reprint,
 superscriptaddress,
 showpacs,preprintnumbers,
nofootinbib,
 amsmath,amssymb,
 aps,
]{revtex4-1}

\usepackage{graphicx}
\usepackage[]{textcomp}
\usepackage[]{xspace}
\usepackage{dcolumn}
\usepackage{color}
\usepackage{color,soul}
\usepackage{graphicx}
\usepackage{dcolumn}
\usepackage{bm}

\begin{document}


\title{Colossal photon bunching in quasiparticle-mediated nanodiamond cathodoluminescence}

\author{Matthew A. Feldman}
\email{Matthew.Feldman@vanderbilt.edu}
\affiliation{Department of Physics and Astronomy, Vanderbilt University, Nashville, Tennessee 37235, USA}
\affiliation{Quantum Information Science Group, Oak Ridge National Laboratory, Oak Ridge, Tennessee 37831, USA}
\author{Eugene F. Dumitrescu}
\affiliation{Quantum Information Science Group, Oak Ridge National Laboratory, Oak Ridge, Tennessee 37831, USA}
\author{Denzel Bridges}
\affiliation{Department of Mechanical Engineering, University of Tennessee, Knoxville, Tennessee 37996, USA}
\author{Matthew F. Chisholm}
\affiliation{Materials Science and Technology Division, Oak Ridge National Laboratory, Oak Ridge, Tennessee 37831, USA}
\author{Roderick B. Davidson}
\affiliation{Department of Physics and Astronomy, Vanderbilt University, Nashville, TN 37235, USA}
\affiliation{Quantum Information Science Group, Oak Ridge National Laboratory, Oak Ridge, Tennessee 37831, USA}
\author{Philip G. Evans}
\affiliation{Quantum Information Science Group, Oak Ridge National Laboratory, Oak Ridge, Tennessee 37831, USA}
\author{Jordan A. Hachtel}
\affiliation{Department of Physics and Astronomy, Vanderbilt University, Nashville, TN 37235, USA}
\affiliation{Center for Nanophase Materials Science, Oak Ridge National Laboratory, Oak Ridge, Tennessee 37831, USA}
\author{Anming Hu}
\affiliation{Department of Mechanical Engineering, University of Tennessee, Knoxville, Tennessee 37996, USA}
\author{Raphael C. Pooser}
\affiliation{Quantum Information Science Group, Oak Ridge National Laboratory, Oak Ridge, Tennessee 37831, USA}
\affiliation{Institute for Functional Imaging of Materials, Oak Ridge National Laboratory, Oak Ridge Tennessee 37831, USA}
\author{Richard F. Haglund}
\affiliation{Department of Physics and Astronomy, Vanderbilt University, Nashville, Tennessee 37235, USA}
\author{Benjamin J. Lawrie}
\email{lawriebj@ornl.gov}
\affiliation{Quantum Information Science Group, Oak Ridge National Laboratory, Oak Ridge, TN 37831, USA}
\affiliation{Institute for Functional Imaging of Materials, Oak Ridge National Laboratory, Oak Ridge TN 37831, USA}

\date{\today}
\begin{abstract}
Nanoscale control over the second-order photon correlation function  $g^{(2)}(\tau)$ is critical to emerging research in nonlinear nanophotonics and integrated quantum information science.  Here we report on quasiparticle control of photon bunching with $g^{(2)}(0)>45$ in the cathodoluminescence of nanodiamond nitrogen vacancy (NV$^0$) centers excited by a converged electron beam in an aberration-corrected scanning transmission electron microscope. Plasmon-mediated NV$^0$ cathodoluminescence exhibits a 16-fold increase in luminescence intensity correlated with a three fold reduction in photon bunching compared with that of uncoupled NV$^0$ centers.  This effect is ascribed to the excitation of single temporally uncorrelated NV$^0$ centers by single surface plasmon polaritons.  Spectrally resolved Hanbury Brown--Twiss interferometry is employed to demonstrate that the bunching is mediated by the NV$^0$ phonon sidebands, while no observable bunching is detected at the zero-phonon line. The data are consistent with fast phonon-mediated recombination dynamics, a conclusion substantiated by agreement between Bayesian regression and Monte Carlo models of superthermal NV$^0$ luminescence. 

\end{abstract}

\maketitle



The efficiency of second-order nonlinearities scales proportionally with $g^{(2)}(0)$, the second-order photon correlation function at zero delay of the driving optical field~\cite{spasibko2017,qu1992}.  Nanoscale superthermal light sources exhibiting photon bunching with $g^{(2)}(0) > 2$ thus provide a path toward high-efficiency nonlinear nanophotonics.  Moreover, control of $g^{(2)}(\tau)$ is increasingly critical for quantum nanophotonics applications~\cite{ridolfo2010quantum,dumitrescu2017zero}. However, despite increasing evidence of coherent quantum behavior in nanoplasmonic systems \cite{lawrie2013extraordinary,tame2013quantum}, experimental plasmonic control of $g^{(2)}(\tau)$ has been realized only in Purcell enhancement of the anti-bunching dynamics of plasmon-coupled emitters~\cite{choy2011enhanced}. 

Compared with photoluminescence (PL) spectroscopy, cathodoluminescence (CL) yields vastly improved spatial resolution in measurements of $g^{(2)}(\tau)$. This fact was leveraged in the first explorations of CL photon statistics, in which photon antibunching was observed from individual NV$^{0}$ centers in nanodiamonds and from point defects in hexagonal boron nitride  excited by an 80-keV electron beam~\cite{tizei2013spatially,bourrellier2016bright,Tizei2017}. More critically, photon bunching has been observed in the CL of ensembles of quantum emitters whose PL exhibits $g^{(2)}(\tau) \approx 1$ because of the absence of temporal correlations between optically excited emitters.  In contrast to PL, the scanning transmission electron microscope (STEM) primarily excites higher-energy modes, such as the 30-eV bulk plasmon in diamond~\cite{zhang2008retrieving}.  The subsequent cascading excitation of multiple excitons and color centers for each plasmon, within an $\sim10$ fs excitation window, explains recent observations of photon bunching of $g^{(2)}(0)-1 > 4$ in CL spectroscopy of ensembles of NV$^{0}$ centers in nanodiamond~\cite{meuret2015photon,Meuret2017}. However, understanding the classical and quantum optical properties of CL generated by semiconducting nanostructures driven by high-energy electron beams requires differentiation between distinct transition pathways for electron- and phonon-mediated luminescence.

In this Rapid Communication, we report observations of room-temperature photon bunching in nanodiamond CL an order of magnitude greater than previously seen at low temperature and more than two orders of magnitude greater than previously seen at equivalent electron-beam currents.  We demonstrate that the bunching is not associated with the NV$^{0}$ zero-phonon line---where we record $g^{(2)}(\tau)-1 \approx 0$---but rather with the phonon sideband. We develop a Monte Carlo model in order to identify the principal physical variables that drive the observed bunching, and compare that model to a Bayesian regression analysis of the measured CL. We also explore photon bunching for ensembles of nanodiamonds evanescently coupled to surface plasmon polaritons (SPPs) supported on a single-crystal Ag nanoplate, recording a 16-fold increase in CL intensity with a concomitant three fold reduction in photon bunching.

Our model suggests that phonon-mediated photon bunching can be attributed to faster recombination dynamics in the phonon sideband compared to bare NV$^{0}$ optical transitions; conversely, reduced photon bunching combined with enhanced CL intensity in the plasmon-coupled composite is consistent with near-resonant SPP excitation of temporally uncorrelated diamond emitters, rather than Purcell-enhanced recombination dynamics.   Taken together, these results point to the possibility of controlling $g^{(2)}(\tau)$ across the visible spectrum with nanoscale spatial resolution by leveraging quasiparticle interactions with the emitters.

\begin{figure}[t!]
    \centering
    \includegraphics[width=0.475\textwidth]{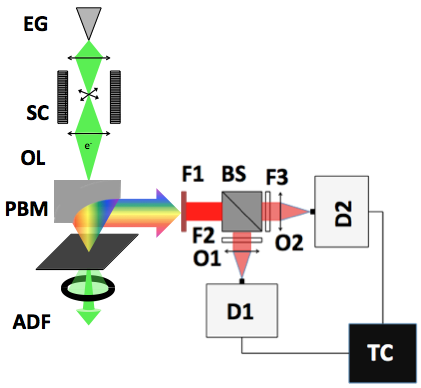}
    \caption{A 2-sr Al parabolic mirror with a pinhole to pass the electron beam was integrated into a VG601 STEM. The collimated CL collected by the parabolic mirror is then characterized by an Acton SP2500 spectrometer or a Hanbury Brown--Twiss interferometer. EG: electron gun; SC: scan coil; OL: objective lens; PBM: parabolic mirror; ADF: annular dark-field detection; F1: zero-phonon line bandpass or phonon sideband long-pass filter; F2,F3: 750 nm short-pass filter; O1,O2: objectives; D1,D2: Perkin Elmer SPCM-AQR; TC: HydraHarp time correlation electronics.}
    \label{fig:fig1}
\end{figure}
Nanodiamonds 120 nm in diameter containing $ \sim1200$  NV$^{0}$ centers per particle were dropcast onto a single-crystal silver nanoplate roughly 100 nm thick and 100 $\mathrm{\mu}$m wide and loaded into an abberation-corrected VG601 STEM.  The STEM, illustrated schematically in Fig.~\ref{fig:fig1}, was operated at room temperature with an electron energy of 60-keV. Cathodoluminescence spectra were acquired in an Acton SP2500 spectrometer.

The second-order correlation function $g^{(2)}(\tau)$ is a normalized measure of photon fluctuations~\cite{glauber1963quantum,carmichael1976quantum} that quantifies the correlation between photons detected at time $t + \tau$ and at time $t$ on two single photon detectors,
{
\begin{align}
g^{(2)}(\tau) &= \frac{\langle \hat{a}^{\dagger}(t)\hat{a}^{\dagger}(t+\tau)\hat{a}(t+\tau)\hat{a}(t) \rangle}{\langle \hat{a}^{\dagger}(t) \hat{a}(t) \rangle \langle \hat{a}^{\dagger}(t+\tau) \hat{a}(t+\tau) \rangle}. 
\end{align}
}Here, $g^{(2)}(\tau)$ of the CL was measured by a Hanbury Brown-Twiss interferometer as shown in Fig.~\ref{fig:fig1}. Photons detected by the single-photon counting modules (SPCMs) were recorded by a HydraHarp 400 time-interval analyzer with 256 ps bin sizes. Infrared photons generated by breakdown flash in the SPCMs~\cite{kurtsiefer2001breakdown} were attenuated by 750 nm short-pass filters. Detected photon pairs were subsequently used to generate $g^{(2)}(\tau)$ statistics for electron beam currents of 0.2--2.1 nA, while power spectra were collected concurrently to confirm the NV$^{0}$ spectrum. Single-photon count rates were 300--10000/s, leading to integration times on the order of an hour and a photon coincidence probability of order $10^{-7}$ per electron.

\begin{figure}[h!]
    \centering
    \includegraphics[width=\linewidth]{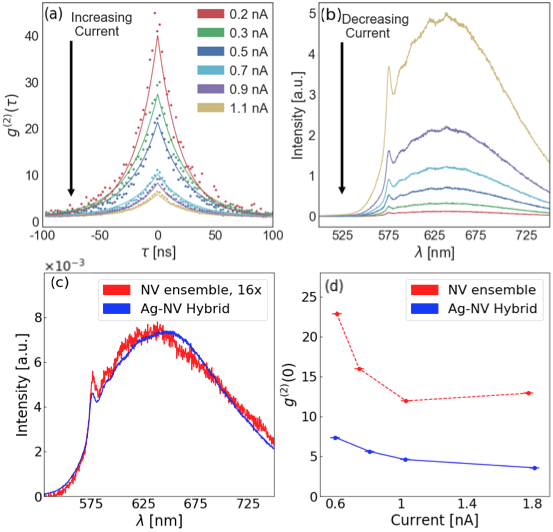}
    \caption{(a) Measured $g^{(2)}(\tau)$ (dots) with Bayesian fits (lines) for diamond-Ag nanocomposite probed using electron beam currents of 0.2--1.1 nA; (b)  CL spectra for the same nanocomposite and currents recorded concurrently with $g^{(2)}(\tau)$. (c) CL spectra illustrating broadband 16-fold increase in CL intensity in the Ag-NV center composite compared with uncoupled NV centers at 0.6 nA. (d) Comparison of median $g^{(2)}(0)$ values for uncoupled nanodiamonds (red) and nanodiamonds evanescently coupled to Ag SPPs. }  
    \label{fig:fig2}
\end{figure}

The $g^{(2)}(\tau)$ of the unfiltered CL---calculated by normalizing the measured photon coincidences to the mean coincidences at $\tau \gg 0$---are plotted in Fig.~\ref{fig:fig2} (a) along with a self-consistent Bayesian regression to a four-parameter exponential decay, as described in the Supplemental Material~\cite{suppmatt, Doherty:2013bu, Sapozhnikov:1976cg, Miyakawa:1970cd}. The standard deviations for the amplitude and the effective lifetime, $\tau_{\text{eff}}$, were less than $5\%$ of their median values. The goodness-of-fit was determined using the mean square error (range: 1.4--6.9) and coefficient of determination (range: 0.88--0.94). Additionally, $1\sigma$, $2\sigma$, and $3\sigma$ credibility intervals and median coincident curves were compared to the data to estimate the precision of our model~\cite{suppmatt}.

Because the statistical distribution of electrons in the beam is Poissonian~\cite{egerton2011electron}, bunching asymptotically approaches $g^{(2)}(0)-1=0$ with increasing beam current. Previous observations of bunching in low-temperature NV$^{0}$ CL reached this limit at a current of 0.1 nA~\cite{meuret2015photon}. As shown in Figs.~\ref{fig:fig2}(a) and 2(d), the measured bunching decreases monotonically with increasing current, but is more than two orders of magnitude greater than bunching previously recorded for these currents at low temperature.

In order to confirm that the bunching was consistent across multiple nanodiamonds, coincident photon counts were measured multiple times to infer the $g^{(2)}(\tau)$ for each electron beam current.  The maximum bunching experimentally observed in 66 measurements was $g^{(2)}(0) = 49.0$ $(0.9)$. The mean fitted lifetime, $\langle \tau_{\text{eff}} \rangle$, across all experiments was $21.1$ $(0.9)$ ns. This is consistent with past reports of NV$^{0}$ lifetimes, which range from 12 to 45 ns depending on the local environment~\cite{liaugaudas2012luminescence,storteboom2015lifetime,meuret2015photon,Doherty:2013bu,inam2013tracking}.  

The CL spectra corresponding to the bunching data shown in Fig.~\ref{fig:fig2}(a) are presented in Fig.~\ref{fig:fig2}(b), with the zero-phonon line well-resolved at a wavelength of 575 nm, and the phonon mediated emission spanning a large bandwidth to the red of 575 nm. Notably, for all currents explored, the broadband transition radiation excited by high energy electrons at dielectric interfaces~\cite{brenny2014quantifying} was unobservable compared with the intensity of the NV$^{0}$ CL. As a result, the contribution of transition radiation to $g^{(2)}(\tau)$ can be assumed to be negligible. The ratio of the intensity of the phonon mediated emission to that of the zero-phonon line and the near-field coupling to Ag SPPs are the only significant differences between the room-temperature CL reported in Fig.~\ref{fig:fig2} and the previous report of low-temperature photon bunching in diamond CL~\cite{meuret2015photon}, but negligible bunching was observed in that report for currents exceeding 0.1 nA.

The effect of plasmonic coupling on the CL second-order coherence function can be explored by comparing $g^{(2)}(\tau)$ for the nanodiamond-Ag hybrid system to $g^{(2)}(\tau)$ for uncoupled nanodiamonds. The CL spectra acquired at 0.6 nA beam current for both systems are overlaid in Fig.~\ref{fig:fig2}(c). Despite no appreciable difference in the CL lineshape, the CL intensity for the hybrid system was a factor of 16 larger. Because of the reduced CL intensity of the bare nanodiamonds compared with the hybrid structures, a defocused electron beam with a waist of $\sim 10$ $\mu m$ was used to excite ensembles of emitters for this comparison~\cite{suppmatt}. Figure~\ref{fig:fig2}(d) clearly shows that the colossal photon bunching is not a result of plasmon-NV center interactions, as the 16-fold increase in CL intensity is correlated with a roughly three fold reduction in $g^{(2)}(0)$ for the four measured electron beam currents. 

The median lifetimes of the plasmon-diamond composite system and the uncoupled nanodiamonds were 22.2 and 21.3 ns, respectively; thus the enhanced CL intensity in Fig.~\ref{fig:fig2}(c) must  be understood as the result of SPP-NV center scattering rather than Purcell enhancement.  Each incident electron excites many SPPs in the 100 nm thick Ag plate, each of which can excite one NV center in any of the nanodiamonds distributed across the Ag plate. This plasmon-NV center scattering enhances the total CL intensity without affecting the recombination rate or the CL linewidth, consistent with the measured results.  Moreover, the random distribution of nanodiamonds across the Ag nanoplate eliminates any temporal coherence in the SPP-excited NV centers, reducing the measured photon bunching.  However, the fabrication of appropriately designed plasmonic nanostructures in which isolated nanodiamonds were coupled to a shared resonant plasmon mode would enable full control over the temporal coherence, and therefore of $g^{(2)}(\tau)$.

The colossal bunching reported here could, in principle, be explained by the superradiance Dicke model ~\cite{jahnke2016giant,hassan1980intensity,dicke1954coherence}, which predicts that the lifetime is inversely proportional to the square of the number of emitters in the driving field. However, no change in lifetime was observed as the electron beam was defocused to encompass three orders of magnitude more emitters in Fig.~\ref{fig:fig2}. Thus, the bunching in Fig.~\ref{fig:fig2} must result from the increased phonon population in room-temperature experiments compared with low-temperature experiments~\cite{meuret2015photon}.

\begin{figure}[]
    \centering
    \includegraphics[width=\columnwidth]{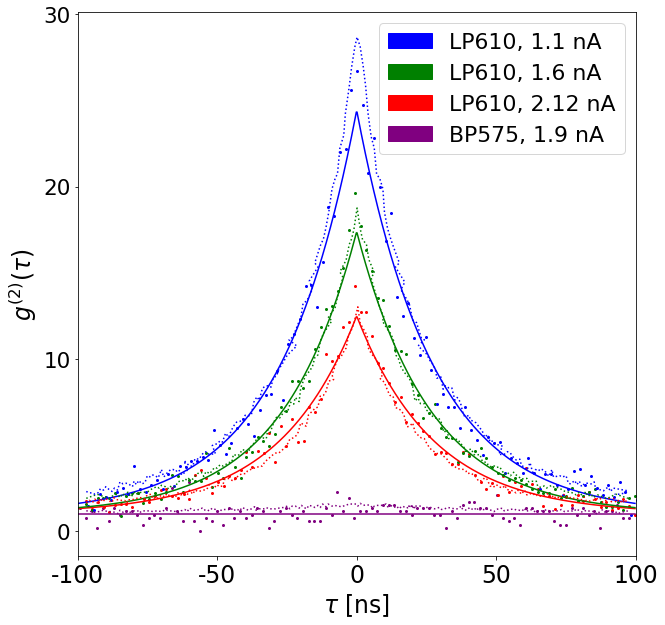}
    \caption{(a) Comparison of measured $g^{(2)}(\tau)$ (dots) to Bayesian median fits (lines) and the Monte Carlo model (dashed lines) for the spectrally (BP575, LP610) filtered CL.} 
    \label{fig:fig3}
\end{figure}

Phononic control was explored by separately filtering the CL of the diamond-Ag composite with a 575-nm bandpass filter with bandwidth of 5 nm and a 610-nm long-pass filter for characterization of the bunching associated with the zero-phonon line and the phonon-mediated CL, respectively. Because of reduced photon counts associated with this spectral filtering, increased electron beam currents of 1.0--2.1 nA were used, although a converged electron beam was used again to address an individual nanodiamond. Figure~\ref{fig:fig3} shows $g^{(2)}(\tau)$ for varying electron beam currents for the 610-nm long-pass filter (LP610), along with $g^{(2)}(\tau)$ for the bandpass filtered CL (BP575) at a current of 1.9 nA. For all currents at which statistically significant coincident counts were measured, no bunching was measured at the zero-phonon line. In contrast, greater bunching was seen for the long-pass filtered CL than for the unfiltered CL at corresponding currents shown in Figs.~\ref{fig:fig2}(a) and ~\ref{fig:fig2}(d). The NV$^-$ color center at 637 nm has never been observed by CL~\cite{Doherty:2013bu}, and does not appear to be present in the spectra shown in Figs.~\ref{fig:fig2}(b) and ~\ref{fig:fig2}(c), but the use of an additional 665-nm long-pass filter ensures that the coincident counts measured are only those of the phonon-mediated emission. While the 665-nm long-pass filter reduced the singles rate to ~600 counts per second, limiting the range of usable currents, the measured photon bunching at a current of 0.9 nA was 11.7 (0.5) compared with the unfiltered CL bunching of 8.3 (0.07) at the same current.

The simplest explanation for this spectral distribution of photon bunching lies in the faster recombination times associated with phonon-mediated luminescence~\cite{storteboom2015lifetime,inam2013tracking}.
A stochastic model incorporating multiple radiative transitions with faster lifetimes for phonon-mediated decay is therefore critical to explaining the orders-of-magnitude increase in photon bunching compared with previous reports \cite{meuret2015photon}. Here, we use a phenomenological Monte-Carlo model of $g^{(2)}(\tau)$ describing the excitation and decay of NV centers  as a function of electron beam current and the lifetimes of multiple phonon-mediated transitions. A full description of the model is provided in the Supplemental Material~\cite{suppmatt,meuret2015photon,Doherty:2013bu, Rothwarf:2003ba,Chapman:2011be,Tizei:2012ck}. The modeled emission time-series data quantitatively reproduce the monotonic decrease of $g^{(2)}(0)$ with respect to electron beam current. More critically, the Monte Carlo model predicts no photon bunching in the zero-phonon line and enhanced bunching in the phonon-mediated CL, as shown in Fig.~\ref{fig:fig3}.

\begin{figure}[]
    \centering
    \includegraphics[width=\columnwidth]{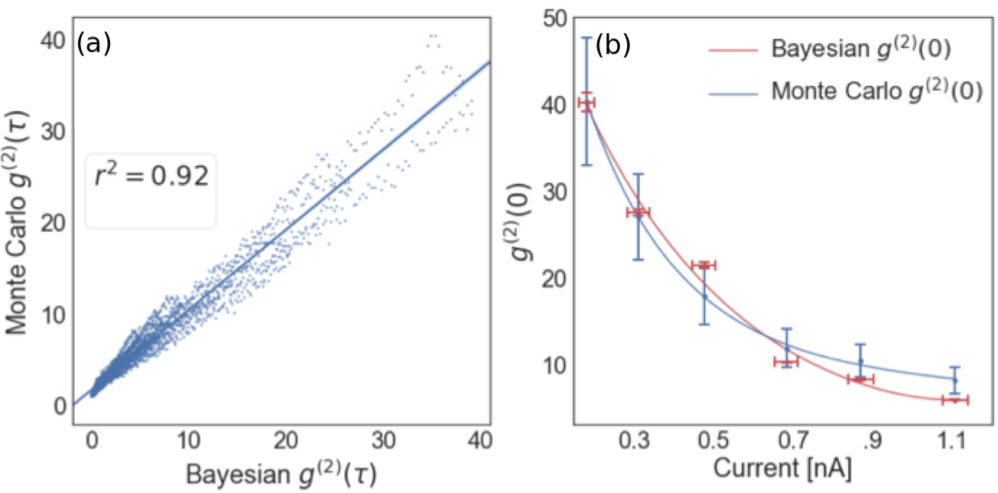}
    \caption{(a) Synthetic Monte Carlo  $g^{(2)}(\tau)$  data as a function of  Bayesian $g^{(2)}(\tau)$ fits to experimental data from Fig.~\ref{fig:fig2}(a) illustrating the quality of the Monte Carlo model. The blue line is a least-squares linear regression. (b) Comparison of the empirical (blue) and synthetic Monte Carlo (red) $g^{(2)}(0)$ data as a function of current.} 
    \label{fig:fig4}
\end{figure}

The Monte Carlo model was validated by determining the correlation between the synthetic data and the Bayesian fits associated with all $g^{(2)}(\tau)$ data from Fig.~\ref{fig:fig2}. The plot of the modeled $g^{(2)}(\tau)$ against the fitted $g^{(2)}(\tau)$ shown in Fig.~\ref{fig:fig4}(a) has a Pearson correlation coefficient of 0.97. A coefficient of determination (0.92) goodness-of-fit test confirmed that a frequentist linear regression produced a reasonable fit. The Monte Carlo simulations also reproduce the observed monotonic decrease in $g^{(2)}(0)$ as a function of electron current as shown in Fig.~\ref{fig:fig4}(b).

The extraordinary variability in the measured $g^{(2)}(\tau)$ across the nanodiamond NV$^0$ CL spectrum, combined with the nanoscale quasiparticle control of $g^{(2)}(\tau)$, points to the need for continuing research into the quantum and classical properties of CL in electron-beam-driven nanomaterials.  The spectrally resolved Monte Carlo simulations presented here provide a mechanistic description of the essential physics in a semiclassical limit. 
However, the nanoscale control of quantum properties of light like $g^{(2)}(\tau)$ in plasmon-emitter nanocomposites is critical to the development of novel applications in quantum information science, including recent proposals for steady-state, dissipatively driven entanglement~\cite{dumitrescu2017zero,gonzalez2011entanglement}.  Microscopic quantum models of electron-beam-driven systems will be required to elucidate the precise connection between the quasiparticle control of $g^{(2)}(\tau)$ shown here and schemes based on driven preparation of specific quantum states.  Notably, these schemes should enable quantum coherent nanoscale control of materials similar to that previously explored with ultrafast electron sources ~\cite{feist2015quantum}, but with more conventional continuous wave electron sources.

While a Ag-diamond nanocomposite was used here to enhance photon count rates, substrate nanopatterning to optimize the Purcell factor of the coupled Ag plasmon-NV center system will enable control over dissipative entanglement schemes and proportional scaling of $g^{(2)}(0)$.  Purcell factors exceeding 1000 have been achieved for optically driven emitters coupled to gold nanocubes~\cite{akselrod2014probing}, but no significant Purcell factors have been reported in electron driven systems \cite{lourenço2017probing}. Optimizing electron driven photon bunching by near-field coupling to plasmonic and dielectric metamaterials with selected phonon-, plasmon-, and substrate-emitter interactions will ultimately provide a critical tool for integrated nonlinear nanophotonics and quantum information science.

\begin{acknowledgments}
This research was sponsored by the Laboratory-Directed Research and Development Program of Oak Ridge National Laboratory, managed by UT-Battelle, LLC for the U.S. Department of Energy. Microscopy studies at the Oak Ridge National Laboratory are supported by the Department of Energy Office of Science, Basic Energy Sciences, Materials Sciences and Engineering Division (MFC). RBD gratefully acknowledges support from the United States Department of Energy, Office of Science (DE-FG02-01ER45916). Matthew Feldman gratefully acknowledges support by the Department of Defense (DoD) through the National Defense Science \& Engineering Graduate Fellowship (NDSEG) Program. The SiN membranes were fabricated at the Center for Nanophase Materials Sciences (CNMS),  which  is  sponsored  at  ORNL  by  the  Scientific  User 
Facilities Division, Office of Basic Energy Sciences, U.S. Department of Energy. We thank Mathieu Kociak and Sophie Meuret for their helpful feedback regarding past cathodoluminescence measurements and Harrison Prosper for his guidance on the Bayesian analysis. 
\end{acknowledgments}

%

\end{document}